\documentstyle[epsfig]{qcdparis}
\newcommand{\wfig}[4]{\begin{figure*}[tb]\vfill\begin{center}
\mbox{\epsfig{figure=#1,height=#2}}\caption{#3}\label{#4}
\end{center}\vfill\end{figure*}}
\newcommand{\Pom}{I$\!$P}

\begin{document}
\pagestyle{plain}
\title{Rapidity Gaps in DIS through Soft Colour Interactions}
 
\author{
\vspace*{-4.3cm}
DESY 95-145 \hfill ISSN 0418-9833\\
July 1995 \hfill \hspace*{1cm}{ }\\
\vspace*{3.1cm}
A.~Edin$^1$, G.~Ingelman$^{1,2}$, J.~Rathsman$^1$}
 
\affil{  
 $^{1)}$Dept. of Radiation Sciences, Uppsala University, 
 Box 535, S-751 21 Uppsala, Sweden\\
 $^{2)}$Deutsches Elektronen-Synchrotron DESY, Notkestrasse 85, D-22603 Hamburg,
 Germany\\}
 
\abstract{We present a new mechanism for the creation of large rapidity gaps
in DIS events  at HERA. Soft colour interactions between perturbatively
produced partons and colour-charges in the proton remnant, modifies the colour
structure for hadronization giving colour singlet systems that are well
separated in rapidity. An explicit model is presented that, although the
detailed results depend on the initial state parton emission, can describe both
the observed  rapidity gaps and, in addition, the forward energy flow in an 
inclusive event sample.}
 
\resume{ \hspace{1cm}}

\twocolumn[\maketitle] 
\fnm{7}{Presented by GI at workshop `DIS and QCD', Paris, April 1995}

A striking feature of deep inelastic scattering (DIS) at the HERA 
$ep$ collider is the relatively large fraction ($\sim 10\%$) of events with 
a rapidity gap \cite{ZEUS,H1}, i.e. with no particles or energy in a large 
rapidity region close to the proton beam direction. These events can be 
interpreted in terms of hard scattering on a pomeron (\Pom ) \cite{IS}, 
a colour singlet object exchanged in a Regge description of diffractive 
interactions. Although explicit such models (see e.g.  \cite{GI,Landshoff}) 
can describe the salient features of the observations, there is no 
satisfactory understanding of the pomeron and its interaction mechanism. 

Here, we present a new and alternative way to interprete the rapidity gap 
phenomenon, without using the concept of a pomeron. Instead, our model is based
on normal DIS parton interactions, with perturbative QCD (pQCD) corrections, 
complemented with the new hypothesis that non-perturbative soft colour 
interactions occur and change the colour structure such that when normal 
hadronization models are applied rapidity gaps may arise. 

At small Bjorken-$x$ ($10^{-4}-10^{-2}$), where the rapidity gap events are
observed,  the boson-gluon-fusion (BGF) process $\gamma g\to q\bar{q}$
(cf.~Fig.~1) occurs  frequently. The cross section is calculable in first order
QCD, with the conventional requirement $m_{ij}^2>y_{cut}W^2$ on any pair $ij$
of partons to avoid soft and collinear divergences. Higher order pQCD emissions
can be taken  into account approximately through parton shower evolution   from
the final partons and the incoming one (as illustrated with one  emitted gluon
in Fig.~1). In the following non-perturbative hadronization  process one
usually considers the formation of colour singlet systems  (clusters, strings)
that subsequently break up into hadrons.  In the conventional Lund model
\cite{Lund} treatment, a BGF event gives  two separate strings from the $q$ and
$\bar{q}$ to the proton remnant spectator partons (Fig.~1a), thereby causing
particle production over the whole rapidity region in between.  This treatment
is implemented in the Monte Carlo LEPTO \cite{LEPTO}, which describes  
most features of HERA DIS events. 
\ffig{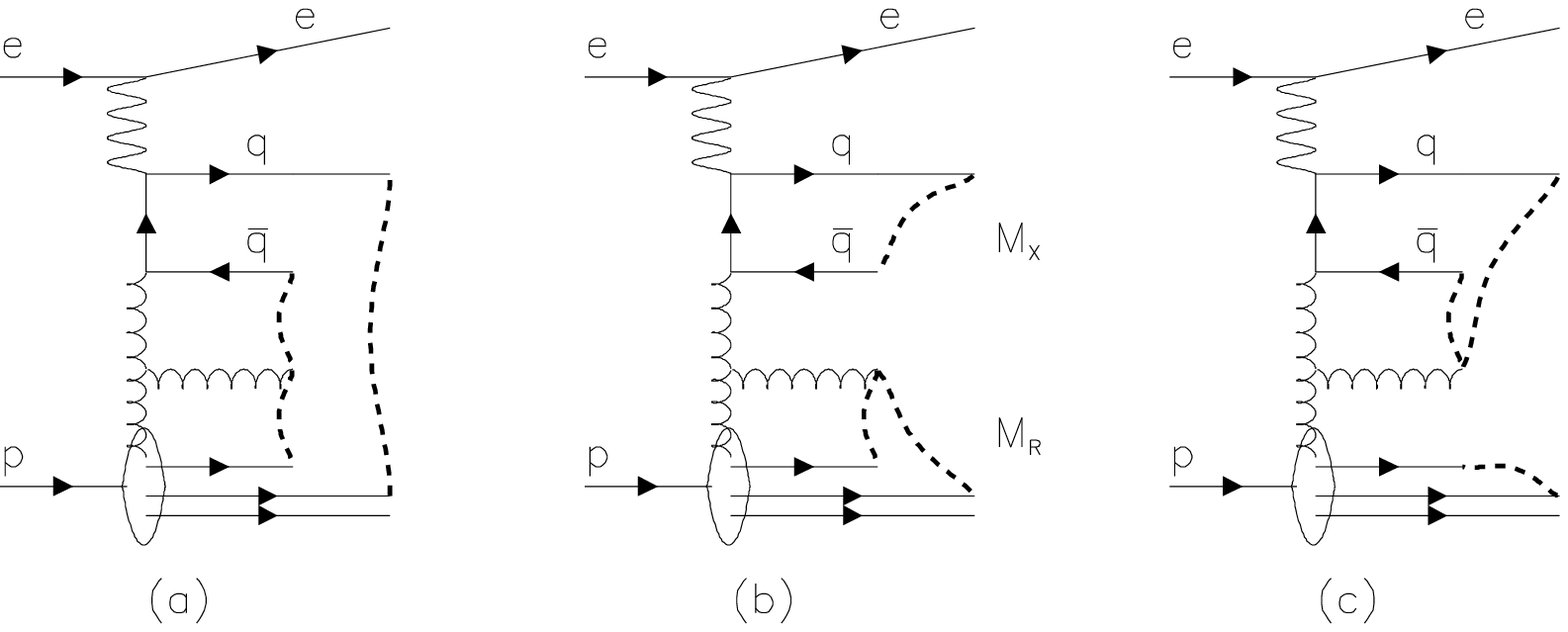}{35mm}{\em The string configuration in a DIS 
boson-gluon-fusion event: 
(a) conventional Lund string connection of partons, and 
(b,c) after reconnection due to soft colour interactions.}{feyn}

This conventional treatment assumes that the colour structure, i.e.\ the 
string topology, follows exactly the perturbative phase with no further 
alternations. Our main assumption here is that additional non-perturbative 
soft colour interactions (SCI) may occur. These have small momentum  transfers,
below the scale $Q_0^2$ defining the limit of pQCD, and do not  significantly
change the momenta from the perturbative phase.  However, SCI will change the
colour of the partons involved and thereby change the colour topology as
represented by the strings. Thus, we propose that the perturbatively produced
quarks and gluons can interact softly with the colour medium of the proton
as they propagate through it. This should be a natural part of the
processes in which `bare'  perturbative partons are `dressed'  into
non-pertubative quarks and gluons and the formation of the confining  colour
flux tube in between them. 

Lacking a proper understanding of such non-perturbative QCD processes, we 
construct a simple model to describe and simulate these interactions.   All
partons from the hard interaction (electroweak $+$ pQCD) plus the  remaining
quarks in the proton remnant constitute a set of colour charges. Each pair of
charges can make a soft interaction changing only the colour and not the
momenta,  which may be viewed as soft non-perturbative gluon exchange.  As the
process is non-perturbative the exchange probability cannot be calculated so
instead we describe it by a parameter $R$.  The number of soft exchanges will
vary event-by-event and change  the colour topology of the events such that, in
some cases, colour singlet subsystems arise separated in rapidity. In the Lund
model  this corresponds to a modified string stretching as illustrated in
Figs.~1bc, where (b) can be seen as a switch of anticolour of the antiquark and
the diquark and (c) as a switch of colour between the two quarks. This kind of
colour switches between the perturbatively produced partons and the valence
partons in the proton are of particular importance for the gap formation.

Of relevance is also the modelling of the non-perturbative proton 
remnant system, i.e. the proton `minus' the parton entering the hard scattering
process. If that parton is a $u$ or $d$ quark, the remnant has previously been
modelled as a valence diquark which, as a colour anti-triplet, will be an 
endpoint of a string. Here we introduce a modified treatment of the remnant
taking into account the possibillity that the quark is a sea quark. 
In that case, the remnant is modelled as three valence quarks, which are
split into a quark and a  diquark in the conventional way  \cite{LEPTO}, 
plus the partner antiquark from the sea to conserve flavour quantum numbers. 
These three partons form two colour singlet systems (strings) 
together with the scattered quark (and extra gluons). This two-string
configuration for sea-quark-initiated processes provides a better continuity to
the two-string gluon-induced BGF events. 
\wfig{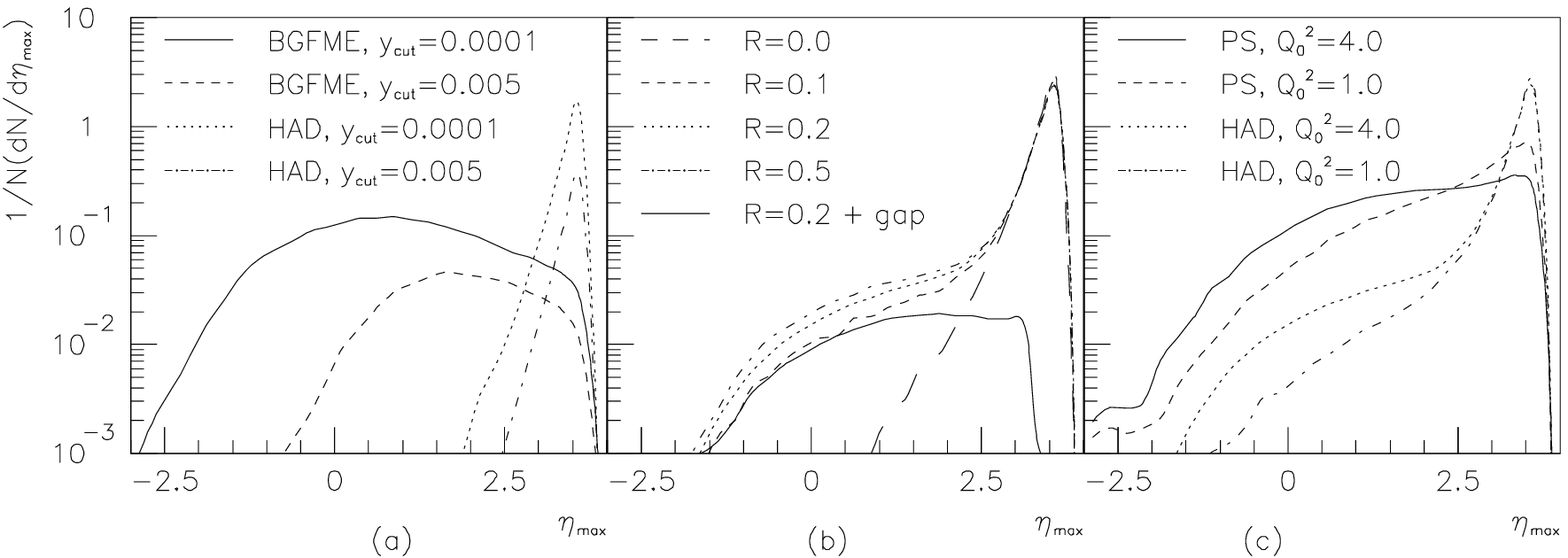}{64mm}{\em Distribution of maximum pseudorapidity 
($\eta_{max}$). (a) Partons from boson-gluon-fusion matrix element (BGFME) 
with cut-off $y_{cut}$, and hadrons (HAD) after parton showers and string 
hadronization (with improved remnant sea quark treatment).  
(b) Hadron level after colour reconnection with probability parameter $R$ for
all events and those satisfying the `gap' definition (full line). 
(c) Parton level (PS) and hadron level (HAD) for different initial state 
parton shower cut-off values $Q_0^2$.}{etamax}

Rapidity gaps have been experimentally investigated \cite{ZEUS,H1} through 
the observable $\eta_{max}$ giving, in each event, the maximum 
pseudo-rapidity in the detector where an energy deposition is present.   
(With $\eta =-\ln{\tan{\theta /2}}$ and $\theta$ the angle relative to 
the proton beam, $\eta > 0$ is the proton hemisphere in the HERA lab frame.) 
Fig.~2 shows the distribution of this quantity as obtained from our model
simulations for $7.5<Q^2<70$ and $0.03<y<0.7$. 

The maximum rapidity parton from the BGF matrix element  (Fig.~2a) can be
central or even in the electron beam hemisphere depending  on the phase space
allowed by $y_{cut}$. For $y_{cut}=0.005$, which has been shown to be
theoretically sound \cite{scale}, about 10 \% are BGF events.
The small $y_{cut}=0.0001$ results \cite{LEPTO} in an adjustment of the 
cut-off such that the total (Born) cross section is saturated with $2+1$-jet 
events, giving $\sim 50\%$ BGF events. 
The partons can here emerge with a large rapidity gap relative to the
spectator partons  at large $\eta$ outside the detector coverage (beam pipe). 
However, this gap does not survive to the hadron level as shown by the  large
effects from parton showering and hadronization.  The introduction of soft
colour interactions have a large effect on the $\eta_{\max}$ distribution, as
demonstrated in Fig.~2b. Still, our SCI model is not very  sensitive to the
exact value of the parameter $R$. In fact, increasing $R$ above $0.5$ gives
almost no effect.  Once a colour exchange with the spectator has occured
additional exchanges  among the partons need not favour gaps and may even
reduce them. In the following we use $R=0.2$, which can be seen as the
strong coupling $\alpha_s(0.5\, GeV)/\pi \approx 0.2$ at a typical small
momentum transfer representative for the region below the perturbative cutoff
$Q_0^2\sim 1\: GeV^2$.  

One should note that the basic features of this distribution, the height of the
peak and the `plateau', is in reasonable agreement with the data \cite{ZEUS,H1}.
(A direct comparison requires taking detailed experimental conditions into 
account.)  Selecting events with rapidity gaps similar to the H1 definition 
(i.e. no energy in $6.6>\eta>\eta_{max}$ where $\eta_{max}<3.2$) 
gives the full curve in Fig.~2b, also in basic agreement with data \cite{H1}.

\wfig{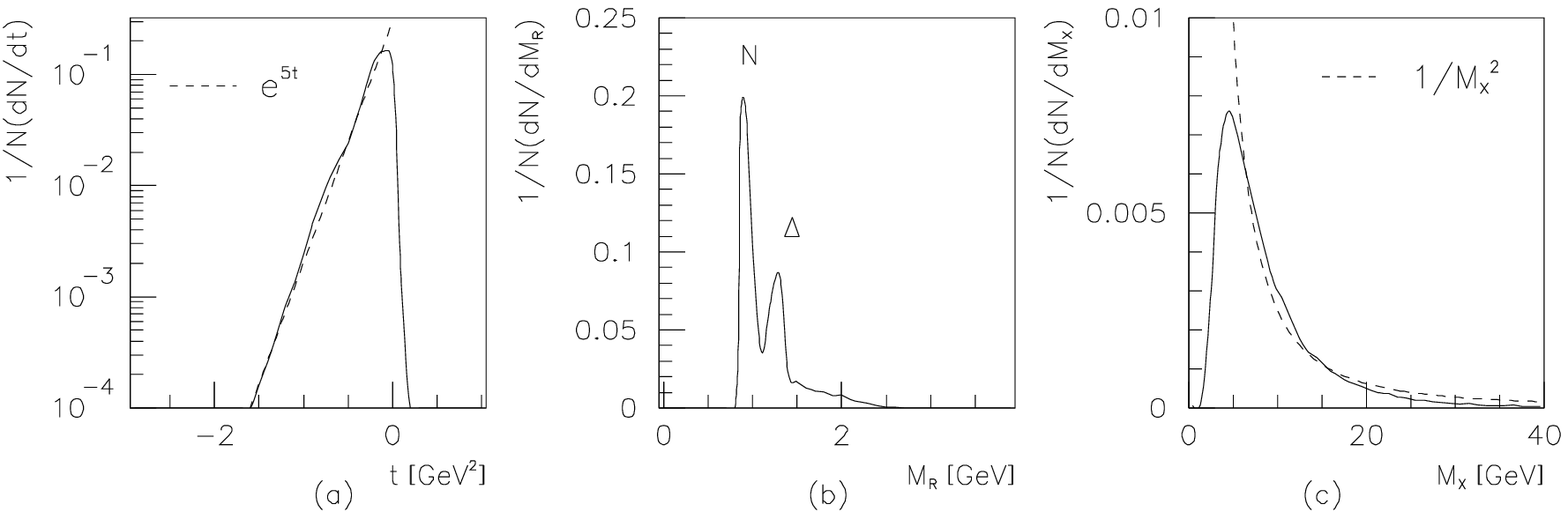}{59mm}{\em Distributions for the selected rapidity gap events, 
i.e. with no energy in $6.6>\eta>\eta_{max}$ and $\eta_{max}<3.2$:
(a) squared momentum transfer $t$ from incoming proton to remnant system $R$;
(b,c) invariant masses of the forward remnant system $M_R$ and of the 
produced central system $M_X$ (cf.~Fig.~1).}{tmxmr}

The exact gap probability does, however, depend on the details of the higher 
order parton emission treatment. In particular, the cut-off $Q_0^2$ for the 
parton shower is rather important. Chosing a value close the hadronic mass
scale $\sim 1\: GeV^2$ tends to produce too much radiation at large rapidity 
such that the gaps are partly destroyed. A value of $4\: GeV^2$ for the limit 
of pQCD, as in many parton density parametrizations, reduces such emissions 
and larger gaps thereby arise after SCI (Fig.~2c). 

In this context one should note that the leading logarithm GLAP evolution need
not be correct when applied to the simulation of exclusive parton final states
in a parton shower. It is derived for not-too-small $x$ and only for the
inclusive case, i.e. for the evolution of the parton density, and sums over all
emissions such that important cancellations can be exploited. It is not clear
whether this formalism is fully applicable also to exclusive final states.
It seems likely that it gives the correct mean behaviour, but it may
not properly estimate the fluctuations that may occur in the emission chain.
Some events may therefore have less parton radiation than estimated in this way
and these would favour the occurence of rapidity gaps. 

Further features of our model are shown in Fig.~3, where the resulting 
distributions in momentum transfer $t$ and mass of the remainder system $R$
and the produced system $X$ (cf.~Fig.~1) are displayed for the selected 
gap events.  
Although the model makes no particular assumptions or requirements on
these quantities, their distributions are similar to those of diffractive 
interactions. This applies to the essentially exponential $t$-dependence, 
$1/M_X^2$ dependence and the $M_R$ system being dominated by the proton with 
some $\Delta$ resonance contribution. Thus, with a gap definition suitable for
selecting diffractive interactions, our model shows the same general 
behaviour as models based on pomeron and other Regge exchanges. However,  
the detailed behaviour depends on the gap definition. Requirements of a large 
gap that extends very forward in rapidity introduces a kinematical bias 
against large values of $t$ and $M_R$.  The $1/M_X^2$ behaviour is explained
by the $1/s_{q\bar{q}}$ dependence of the BGF matrix elements, 
but is distorted at large $M_X$ by requiring the gap to extend into
the central rapidity region. 
Thus, by varying the gap definition or observing the 
forward-moving $R$-system, one may find observable differences between the 
two kinds of models. 

The rate of gap events in the model is essentially independent of the 
DIS kinematical variables $x$ and $Q^2$. This is a natural consequence 
of the assumed factorization of the SCI relative to the hard perturbative 
interactions. Some small variation may still occur since the parton shower
details depends on $x,Q^2$. 

\ffig{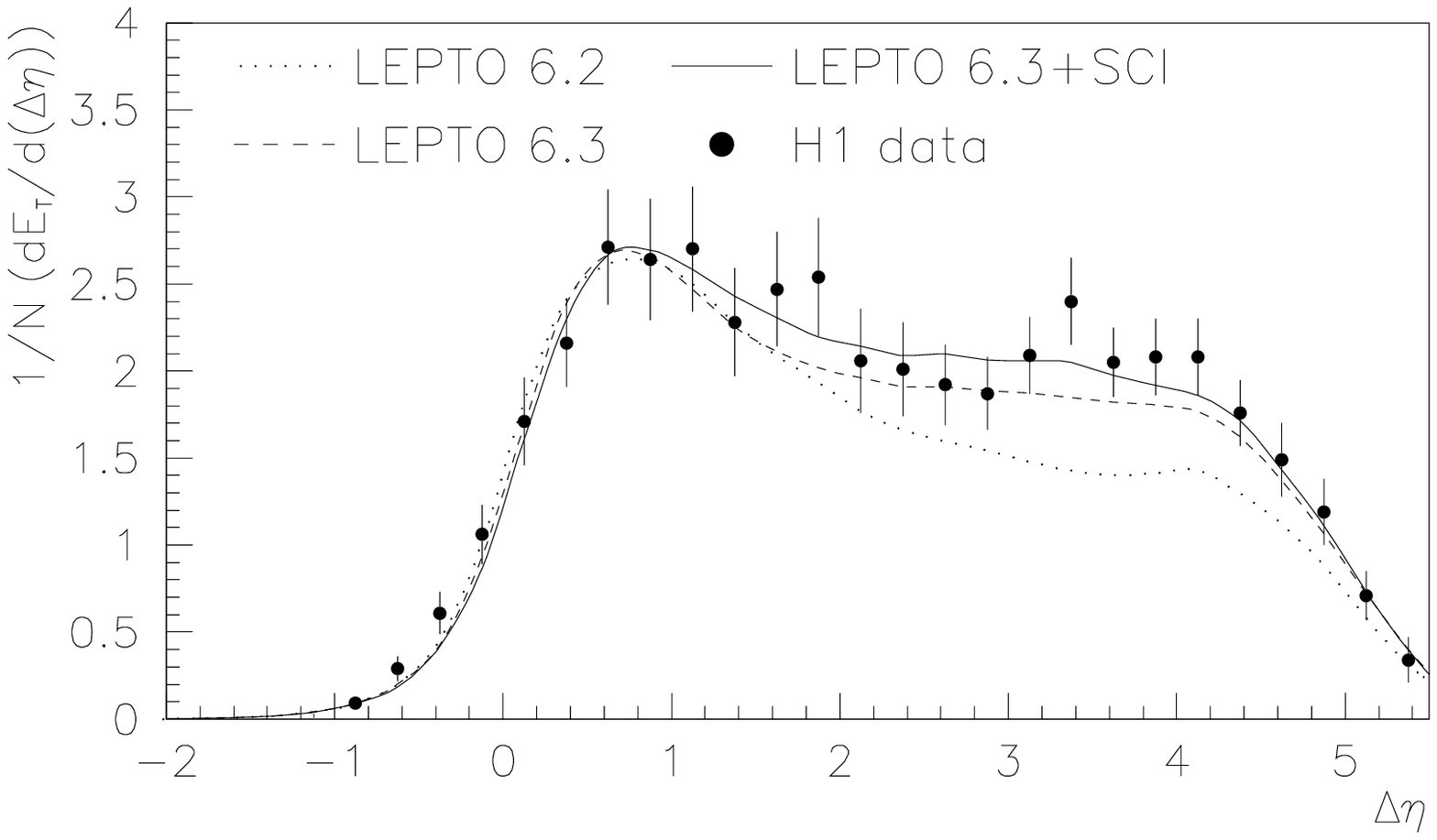}{50mm}{\em Transverse energy flow versus 
$\Delta\eta =\eta-\eta_{q/QPM}$, i.e. lab pseudorapidity relative to the 
current direction in QPM kinematics, for events with $x<10^{-3}$.  
The curves are from the earlier Monte Carlo model (6.2), 
with improved parton shower and sea quark treatment (6.3) 
and, in addition, with soft colour interactions (6.3+SCI). }{eflow}

The observed quantity $F_2^D(\beta ,Q^2)$ \cite{H1} can been interpreted as 
the pomeron $F_2$ structure function \cite{IP} 
with conventional QCD evolution 
of partons with momentum fraction $\beta$ in the pomeron \cite{POMevol}. 
Having no pomeron in our model, one should instead consider the evolution of 
partons with momentum fraction $x$ in the proton. Data on $F_2^D$ 
provides another detailed testing ground for the models. 

An observable which gives complementary information relative to the rapidity 
gaps is the forward transverse energy flow. Whereas substantial initial state 
parton radiation spoils the rapidity gaps, it helps to describe the high level
of the forward energy flow. It is therefore a highly non-trivial test of any 
model that both these observables can be accounted for.
As shown in Fig.~4, the $E_T$-flow data \cite{H1eflow} can be well described
by our new model. The two-string configurations from
the improved sea quark treatment contributes to enhancing the energy flow. 
This is also the case for the SCI, which can lead to configurations where 
the string goes `back and forth' producing more energy per unit rapidity.

In conclusion, we have suggested a new mechanism for the production of events
with rapidity gaps. The basic assumption is that soft colour interactions may
occur after the short space-time pertubative phase. This modifies the colour
structure for hadronisation giving colour singlet systems that are separated
in rapidity. A detailed model has features that are characteristic
for diffractive scattering and can qualitatively account for the rapidity gap
events observed in DIS at HERA.

We are grateful to W.~Buchm\"uller for interesting discussions 
and to the Paris workshop organizers for a stimulating meeting.  


\Bibliography{100}
%
\bibitem{ZEUS}
 ZEUS Coll., Phys.~Lett.~B315~(1993)~481; DESY 95-093
\bibitem{H1} 
 H1 Collaboration, Nucl. Phys. B429 (1994) 477; Phys. Lett. B348 (1995) 681
\bibitem{IS}
 G.~Ingelman, P.E.~Schlein, Phys. Lett. B152 (1985) 256
\bibitem{GI}
 G.~Ingelman, J.~Phys.~G:~Nucl.~Part.~Phys.~19~(1993)~1633
\bibitem{Landshoff}
 P.V.~Landshoff, these proceedings
\bibitem{Lund}
 B.~Andersson et al., Phys. Rep. 97 (1983) 31
\bibitem{LEPTO}
 G.~Ingelman, proc.~`Physics at HERA', DESY 1991, p.~1366
\bibitem{scale}
 G.~Ingelman, J.~Rathsman, Z.~Phys.~C63~(1994)~589
\bibitem{IP} 
 G.~Ingelman, K.~Prytz, Z.~Phys.~C58~(1993)~285
\bibitem{POMevol} 
 See contributions by H.G.~Kohrs, A.~Kaidalov, and  
 K.~Golec-Biernat in these proceedings
\bibitem{H1eflow}
 H1 Collaboration, I.~Abt et al., Z.~Phys.~C63~(1994)~377
\end{thebibliography}
\end{document}